\documentclass[11pt]{article}
\usepackage{graphicx}
\usepackage{amsmath,amsfonts,amsthm}
\usepackage{amssymb,latexsym,mathabx}
\usepackage{fixmath}
\usepackage{mathrsfs,amsbsy}
\usepackage{dsfont}
\newtheorem{theorem}{Theorem}[section]

\newtheorem{proposition}[theorem]{Proposition}
\newtheorem{corollary}[theorem]{Corollary}

\newtheorem{lemma}[theorem]{Lemma}
\newtheorem{example}[theorem]{Example}

\newtheorem{algorithm}[theorem]{Algorithm}
\newtheorem{remark}[theorem]{Remark}

\numberwithin{equation}{section}

\def\IR{{\bf R}}
\def\IC{{\bf C}}

\def\diag{{\rm diag}\,}
\def\rank{{\rm rank}\,}
\def\tr{{\rm tr}\,}

\def\vec{{\rm vec}}

\def\qed{\hfill $\Box$\medskip}
\def\diag{{\rm diag}\,}

\def\IR{{\bf R}}
\def\IC{{\bf C}}

\def\cS{{\cal S}}

\def\tr{{\rm tr}}
\def\a{{\bf a}}
\def\b{{\bf b}}
\def\c{{\bf c}}

\def\IC{{\mathbb C}}
\def\IR{{\mathbb R}}

\def\cH{{\mathcal H}}

\def\cH{{\mathcal H}}

\def\cD{{\mathcal D}}
\def\cM{{\mathcal M}}

\def\cS{{\mathcal S}}
\def\cU{{\mathcal U}}

\def\diag{{\rm diag}\,}

\def\tr{{\rm tr}}

\def\rank{{\rm rank}\,}
\def\diag{{\rm diag}}
\def\[{\left [}
\def\]{\right ]}
\def\({\left (}
\def\){\right )}

\def\Ra{{\ \Rightarrow\ }}

\def\Lra{{\ \Leftrightarrow\ }}
\def\dfrac{\displaystyle\frac}

\begin{document}
\baselineskip 15.5pt
\title{Ranks of quantum states with prescribed reduced states}
\date{}
\author{Chi-Kwong Li, Yiu-Tung Poon, and Xuefeng Wang}
\maketitle
\begin{center}

DEPARTMENT OF MATHEMATICS, COLLEGE OF WILLIAM AND MARY,
\\
WILLIAMSBURG, VA 23187, USA. (CKLI@MATH.WM.EDU)

DEPARTMENT OF MATHEMATICS, IOWA STATE UNIVERSITY,
\\
AMES, IA 50011, USA. (YTPOON@IASTATE.EDU)

SCHOOL OF MATHEMATICAL SCIENCES, OCEAN UNIVERSITY OF CHINA,
\\
QINGDAO, SHANDONG 266100, CHINA. (WANGXUEFENG@OUC.EDU.CN)

\end{center}

\begin{abstract}
Let $\cM_n$ be the set of $n\times n$ complex matrices.
In this note we determine all the possible ranks of a bipartite state
in $\cM_m\otimes \cM_n$ with
prescribed reduced states in the two subsystems. The results are used to
determine the Choi rank of quantum channels $\Phi: \cM_m \rightarrow
\cM_n$ sending $I/m$ to a specific state $\sigma_2 \in \cM_n$.
\end{abstract}

Keywords. Quantum state, reduced state, partial trace, quantum channel, rank, eigenvalue.

AMS Classification. 15A18, 15A60, 15A42, 15B48, 46N50.

\section{Introduction}

In quantum information science, quantum states are used to store,
process, and transmit information. Mathematically,  quantum states
are represented by density matrices, i.e., positive semidefinite
matrices of trace 1.

Let $\cM_n$ ($\cH_n$) be the set of $n\times n$ complex (Hermitian) matrices.
Let  $\cD_n$ be
the set of density matrices in $\cH_n$.
Suppose $\sigma_1 \in \cD_m$ and $\sigma_2 \in \cD_n$ are two quantum states.
Their product state is $\sigma_1 \otimes \sigma_2 \in \cD_{mn}$.
The combined system is known as the bipartite system,
and a general quantum state is represented by a density matrix $\rho \in \cD_{mn}$.

Two important quantum operations used to extract information of the subsystems
from a quantum state of the bipartite system are the partial traces defined as
$$\tr_1(\sigma_1\otimes \sigma_2) = \sigma_2 \quad \hbox{ and } \quad
\tr_2(\sigma_1\otimes \sigma_2) = \sigma_1$$
on tensor states $\sigma_1 \otimes \sigma_2 \in \cD_{mn}$,
and extended by linearity for general states in $\cD_{mn}$.
In particular, suppose $\rho = (\rho_{ij})_{1 \le i, j \le m} \in \cD_{mn}$
such that $\rho_{ij} \in \cM_n$. Then
$$\tr_1(\rho) = \rho_{11} + \cdots + \rho_{mm} \in \cM_n \quad \hbox{ and }
\quad
\tr_2(\rho) = (\tr \rho_{ij})_{1 \le i, j \le m} \in \cM_m.$$

Let $\rank(\sigma)$ be the rank of a matrix $\sigma$.
The purpose of this note is to give a complete answer to the following.

\medskip\noindent
{\bf Problem} Determine all the possible values of
$\rank(\rho)$ for $\rho \in \cD_{mn}$ in the set
$$\cS(\sigma_1, \sigma_2) = \{\rho \in \cD_{mn}: \tr_1(\rho) = \sigma_2,
\tr_2(\rho) = \sigma_1\}$$
for given $\sigma_1 \in \cD_m$ and $\sigma_2 \in \cD_n$.

\medskip
It is well known and easy to verify that the maximum rank of $\rho \in \cS(\sigma_1,\sigma_2)$
equals  $R = \rank(\sigma_1)\rank(\sigma_2)$.
In Section 2, we will present a finite algorithm for determining the minimum
rank value $r$ of $\rho \in \cS(\sigma_1, \sigma_2)$ in terms of the eigenvalues
of $\sigma_1$ and $\sigma_2$. Moreover, we will show that there is $\rho \in
\cS(\sigma_1,\sigma_2)$ with $\rank(\rho) = k$ for every $k \in\{r, r+1, \dots, R\}$.

\medskip
In section 3, we will describe some implications of the results in Section 2 to
the study of quantum channels. In particular, the results allow us to determine all
the possible Choi ranks of a quantum channel 
$\Phi:\cM_m \rightarrow \cM_n$ having a prescribed state
$\Phi(I_m/m)\in \cD_n$.

\medskip
In our discussion,
let $\cM_{m,n}$ be the set of $m\times n$ complex matrices
so that $\cM_n = \cM_{n,n}$.
Denote by $\cU_n$ and $\cH_n$ the set of unitary matrices and the set of
Hermitian matrices in $\cM_n$, respectively.
Let $\IR^n_{\downarrow}$ denote the set of $n$-tuples in $\IR^n$ with decreasing coordinates.
Given $\sigma \in \cH_n$,
$\lambda(\sigma)=(\lambda_1(\sigma),\dots,\lambda_n(\sigma))\in \IR^n_{\downarrow}$ will
denote the eigenvalues of $\sigma$ arranged in decreasing order.

Let
$\{e_1^{(m)}, \dots, e_m^{(m)}\}$ and
$\{e_1^{(n)}, \dots, e_n^{(n)}\}$ be the standard bases
for $\IC^m$ and $\IC^n$, respectively. Then, clearly,
$\{e_1^{(m)}\otimes e_1^{(n)},
e_1^{(m)}\otimes e_2^{(n)}, \dots, e_m^{(m)}
\otimes e_n^{(n)}\}$ is the standard basis for
$\IC^{m}\otimes \IC^n \equiv \IC^{mn}$.
For simplicity, we use the notation $e_i$ for $e_i^{(m)}$ or $e_i^{(n)}$
if the dimension of the vector is clear in the context.
Also, we use $e_i\otimes e_j$ instead of $e_i^{(m)}\otimes e_j^{(n)}$,
and $e_i e_j^t = E_{ij}$ to denote the basic complex unit of appropriate size.

The following observation is useful in our discussion.

\begin{lemma}\label{1.1}
Let $\sigma_1 \in \cD_m$, $\sigma_2 \in \cD_n$, $U \in \cM_m$  and $V \in \cM_n$.
Then
$$\cS(U\sigma_1 U^*,V\sigma_2V^*) =  (U\otimes V) \cS(\sigma_1,\sigma_2) (U\otimes V)^*
= \{ (U\otimes V)\rho (U\otimes V)^*: \rho \in \cS(\sigma_1,\sigma_2)\}.
$$
\end{lemma}

\section{Ranks of matrices in $\cS(\sigma_1, \sigma_2)$}

In this section, we present results concerning the ranks of states in
$\cS(\sigma_1,\sigma_2)$. 
We will use the fact that $\cS(\sigma_1, \sigma_2)$ contains
a rank one matrix if and only if 
$\sigma_1$ and $\sigma_2$ 
have the same nonzero eigenvalues counting multiplicities;
see \cite{Fu,Kl2}.

For $w = (w_1, \dots, w_{mn})^t \in \IC^{mn}$,
let $W = [w]$ be the $m\times n$ matrix such that the $i$th row equals
$(w_{(i-1)n+1}, \dots, w_{in})$ for $i = 1, \dots, m$.
Suppose $W=[w]$ has singular value decomposition $XSY^t$ such that $X \in \cM_m$
is unitary with columns
$x_1, \dots, x_m$, $Y \in \cM_n$ is unitary with columns $y_1, \dots, y_n$ and
$S = s_{1} e_1e_1^t + \cdots + s_k e_k e_k^t$, where $k$ is the rank of $W$.
It follows that $W = [w] = \sum_{j=1}^k s_j x_j y_j^t$, and
$w = \sum_{j=1}^k s_j x_j \otimes y_j$,
which is known as the Schmidt decomposition of $w$.

\begin{theorem} \label{thm2.4} Let $(\sigma_1, \sigma_2) \in \cD_m\times \cD_n$
such that $\rank(\sigma_1) = r_1 \ge r_2 = \rank(\sigma_2)$.
There is an element in $\cS(\sigma_1,\sigma_2)$ attaining the
lowest rank $r$ with $r \le r_1$. Moreover,
there exists $\rho \in  \cS(\sigma_1,\sigma_2)$ of rank $k$ if and only if
$r \le k \le r_1r_2$.
\end{theorem}

\it Proof. \rm By Lemma \ref{1.1}, we may assume that

\medskip\centerline{
$\sigma_1 = \diag(\mu_1,\dots, \mu_{r_1}, 0, \dots, 0)$
and $\sigma_2 = \diag(\hat \mu_1, \dots, \hat \mu_{r_2}, 0, \dots, 0)$.
}

\medskip\noindent
If $\rho = \sum_{j=1}^k z_j z_j^* \in \cS(\sigma_1,\sigma_2)$,
then $\tr_1(\rho) = \sigma_2$ and $\tr_2(\rho) = \sigma_1$.

For $1 \le i \le r_1$ and $1 \le j \le r_2$,
let $e_i\otimes e_j$ denote $e_i^{(m)} \otimes e_j^{(n)}$, and
let
$v_{ij} =  \sqrt{\mu_{i}\hat \mu_{j}}
e_i \otimes e_{j}$.
Now, for $\ell = 1, \dots, r_1$, let
$$z_{\ell} =\left\{ \begin{array}{ll}
\sum_{1 \le i \le r_2}  v_{i+\ell-1,i}&\mbox{ if }1\le \ell\le r_1-r_2+1
\\&\\
\sum_{1 \le i \le r_1+1-\ell}  v_{i+\ell-1,i}
+ \sum_{r_1+1-\ell < i \le r_2} v_{i+\ell-1-r_1,i}&\mbox{ if } r_1-r_2+1< \ell\le r_1\,.
\end{array}
\right.
$$
For example, $z_1 = v_{11} + \cdots + v_{r_2,r_2}$,
$z_2 = v_{21} + v_{32} + \cdots +v_{r_2+1,r_2},\dots, z_{r_1} = v_{r_1,1} + v_{12} + \cdots +v_{r_2-1,r_2}$.
Then

\begin{equation}\label{tr1}
\tr_1\(z_{\ell}z_{\ell}^*\)=\left\{ \begin{array}{ll}
\sum_{1 \le i \le r_2}  \mu_{i+\ell-1}\hat\mu_iE_{ii} &\mbox{ if }1\le \ell\le r_1-r_2+1
\\&\\
\begin{array}{l}\sum_{1 \le i \le r_1+1-\ell}  \mu_{i+\ell-1}\hat\mu_iE_{ii} \\
+ \sum_{r_1+1-\ell < i \le r_2} \mu_{i+\ell-1-r_1}\hat\mu_iE_{ii} \end{array}&\mbox{ if } r_1-r_2+1< \ell\le r_1
\end{array}
\right.
\end{equation}
and 

\begin{equation}\label{tr2}
\tr_2\(z_{\ell}z_{\ell}^*\)=\left\{ \begin{array}{ll}
\sum_{1 \le i \le r_2}  \mu_{i+\ell-1}\hat\mu_iE_{i+\ell-1\ i+\ell-1 }&\mbox{ if }1\le \ell\le r_1-r_2+1
\\&\\
\begin{array}{l} \sum_{1 \le i \le r_1+1-\ell}  \mu_{i+\ell-1}\hat\mu_iE_{i+\ell-1\ i+\ell-1 }\\
+  \sum_{r_1+1-\ell < i \le r_2} \mu_{i+\ell-1-r_1}\hat\mu_iE_{i+\ell-1-r_1\ i+\ell-1-r_1}\end{array}
&\mbox{ if } r_1-r_2+1< \ell\le r_1\,.
\end{array}
\right.
\end{equation}

Let

\begin{equation}\label{rho}
\rho = \sum_{\ell=1}^{r_1} z_\ell z_\ell^* \in \cS(\sigma_1,\sigma_2)\,.
\end{equation}
Then $\rho $ has rank $r_1$. It follows from (\ref{tr1}) and (\ref{tr2}) that $ \rho \in \cS(\sigma_1,\sigma_2)$.
In (\ref{rho}) if we replace
$z_1z_1^*$ by
$$v_{11}v_{11}^* + \cdots + v_{pp}v_{pp}^* +
(\sum_{j>p} v_{jj})(\sum_{j>p} v_{jj})^*, \quad p = 1, \dots, r_2-1,$$
the resulting state is in $\cS(\sigma_1,\sigma_2)$ with rank
$p+r_1$ for $p = 1, \dots, r_2-1$.
Similarly, we can replace each $z_jz_j^*$ by a rank $p+1$ matrix
for $p = 1, \dots, r_2-1$, in such a way that the resulting state still lies
in $\cS(\sigma_1, \sigma_2)$. Hence, $\cS(\sigma_1,\sigma_2)$
contains matrices of rank $k$ for every $k \in \{r_1, \dots, r_1r_2\}$.

Now, suppose $\rho\in \cS(\sigma_1, \sigma_2)$ has rank $r < r_1$.
We will show that there exists $\tilde \rho \in \cS(\sigma_1,\sigma_2)$
with rank $k$ for any $r < k < r_1$.
We prove the result by induction on $r$.
To this end, let
\begin{equation}\label{rho2}
\rho = \sum_{j=1}^{r} z_j z_j^* \in \cS(\sigma_1,\sigma_2).
\end{equation}
By the Schmidt decomposition,  
$z_j = \sum_{\ell = 1}^{t_j} s_{j\ell} x_{j\ell}\otimes y_{j\ell}$
for each $j$, where $t_j = \rank([z_j])$.
Let $w_{j\ell} = s_{j\ell} x_{j\ell}\otimes y_{j\ell}$.
Similar to the previous case, we can replace $z_jz_j^*$
by
$$\sum_{\ell \le p} w_{j\ell}w_{j\ell}^* +
\(\sum_{\ell > p} w_{j\ell}\)\(\sum_{\ell > p}   w_{j\ell}\)^*$$
so that the resulting state still lies in $\cS(\sigma_1,\sigma_2)$.
We may increase the number of summands in (\ref{rho2}) by one at each time
until we get $\hat \rho = \sum_{j,\ell} w_{j\ell}w_{j\ell}^*$.
Note that in each step, the rank of the state will either stay the same
or increase by 1, and
$\rank(\hat\rho) \ge \rank( \tr_2(\hat \rho)) = \rank(\sigma_1) = r_1$.
As a result, the set $\cS(\sigma_1,\sigma_2)$
contains matrices of ranks $r, \dots, r_1$.
\qed

We illustrate the above theorem by the following example.

\begin{example}  \rm Suppose
$\sigma_1 = \diag(a_1, a_2, a_3)$ and $\sigma_2 = \diag(b_1, b_2)$
with $a_1 \ge a_2 \ge a_3 > 0, b_1 \ge b_2 > 0$.
Let $z_1 = (\sqrt{a_1b_1}, 0, 0, \sqrt{a_2b_2}, 0, 0)^t$,
$z_2 = (0, 0, \sqrt{a_2 b_1}, 0,  0, \sqrt{a_3b_2})^t$,
$z_3 = (0, \sqrt{a_1b_2}, 0,0, \sqrt{a_3b_1} ,0)^t$.
Then $\rho = z_1z_1^* + z_2z_2^* + z_3 z_3^* \in \cS(\sigma_1, \sigma_2)$ is of
rank 3. One can replace $z_1z_1^*$ by $v_1v_1^* + v_2 v_2^*$
with $v_1 = (\sqrt{a_1b_1}, 0,0,0,0,0)^t$ and $v_2 = (0,0,0, \sqrt{a_2b_2}, 0, 0)^t$
to get a rank 4 matrix in $\cS(\sigma_1, \sigma_2)$. Similarly, 
we can further replace
$z_2z_2^*$ by $v_3v_3^* + v_4v_4^*$, and $z_3z_3^*$ by $v_5v_5^* + v_6v_6^*$, etc. 
to get matrices in $\cS(\sigma_1, \sigma_2)$ of rank $5$ and $6$.
\end{example}

\begin{corollary} \label{cor2.3} Let $(\sigma_1, \sigma_2) \in \cD_m\times \cD_n$.
There is a rank one $\rho \in \cS(\sigma_1,\sigma_2)$ if and only if
$\sigma_1$ and $\sigma_2$ have the same (multi-)set of non-zero eigenvalues,
say, $\lambda_1 \ge \cdots \ge \lambda_r$. 
 
In such a case,
there exists $\rho \in  \cS(\sigma_1,\sigma_2)$ of rank $k$ if and only if
$1 \le k \le r^2$.
 \end{corollary}

\it Proof. \rm The first part follows from the fact that $\rho = vv^* \in \cS(\sigma_1,\sigma_2)$, where
$v$ has Schmidt decomposition $\sum_{j=1}^r s_j x_j \otimes y_j$,
if and only if $\sigma_1 = \sum_{j=1}^r s_j^2 x_jx_j^*$ and
$\sigma_2 = \sum_{j=1}^r s_j^2 y_jy_j^*$.

The second part follows from Theorem \ref{thm2.4}.
\qed

We illustrate the above Corollary by the following example.

\begin{example}\rm  Suppose $\sigma_1 = \diag(\lambda_1, \lambda_2,0)$ 
and $\sigma_2 = \diag(\lambda_1,\lambda_2)$
such that $\lambda_1 \ge \lambda_2 > 0$ and $\lambda_1 + \lambda_2 = 1$. 
Let 
$f_1 = (\sqrt{\lambda_1}, 0, 0, 0, 0, 0)^t, f_2 = (0,0,0,\sqrt{\lambda_2},0,0)^t$.
Then $(f_1+f_2)(f_1+f_2)^* \in \cS(\sigma_1, \sigma_2)$ has rank 1,
and $f_1f_1^* + f_2f_2^* \in \cS(\sigma_1, \sigma_2)$ has rank 2.
Let $$v_{11} = (\lambda_1, 0, 0,0, 0, 0)^t , 
 v_{12} = (0, 0, 0,\lambda_2, 0, 0)^t , 
 v_{21} = (0, 0, \sqrt{\lambda_2 \lambda_1}, 0,  0,0)^t ,
 v_{31} = (0, \sqrt{\lambda_1\lambda_2}, 0,0, 0 ,0)^t.$$ Then 
$(v_{11}+v_{12})(v_{11}+v_{12})^*+v_{21}v_{21}^*+v_{31}v_{31}^* 
\in \cS(\sigma_1, \sigma_2)$ has rank 3 and 
  $v_{11}v_{11}^* + v_{12}v_{12}^* + +v_{21}v_{21}^*+v_{31}v_{31}^* 
\in \cS(\sigma_1, \sigma_2)$ has rank 4.
\end{example}

Next, we determine the minimal rank of $\rho$ in $\cS(\sigma_1, \sigma_2)$.
For $w = (w_1, \dots, w_{mn})^t \in \IC^{mn}$, we continue to
let $W = [w]$ be the $m\times n$ matrix such that the $i$th row equals
$(w_{(i-1)n+1}, \dots, w_{in})$ for $i = 1, \dots, m$.
One can easily construct the inverse map which converts
an $m\times n$ matrix $W$ to $w = \vec(W) \in \IC^{mn}$
so that $W = [w]$.
Note that $\tr_1(ww^*) = W^t(W^t)^*$ and $\tr_2(ww^*) = WW^*$.
Suppose $\rho\in \cS(\sigma_1, \sigma_2)$ has rank $\le r$. Then there exists
an $mn \times r$ matrix $V$ such that $\tr_1\(VV^*\)=\sigma_2$ and  $\tr_2\(VV^*\)=\sigma_1$.
For $1\le j\le r$,  let $W_j=[v_j]$, where $v_j$ is the j${}^{\rm th}$ column of $V$.
Then we have
 $\sigma_1=W_1(W_1)^*+\cdots+W_r(W_r)^*$ and $\sigma_2=W^t_1\(W_1^t\)^*+\cdots+W^t_r\(W_r^t\)^*$.

Given $C\in H_m$, let $\lambda(C)=(c_1,\dots,c_m)$ denote the eigenvalues of $C$ with $c_1\ge\cdots\ge c_m$. Let $\IR^m_{\downarrow}=\{(x_i)\in \IR^m:x_1\ge\cdots\ge x_m\}$. By a result of Klyachko \cite{Kl1,Fu}, the eigenvalues of a sum of Hermitian matrices can be characterized by    a set  $LR(m,r)$ of $r+1$ tuples
$(J_0,J_1, \dots, J_r)$, where
$J_0, \dots, J_r$ are subsets of $\{1, \dots, m\}$ . More specifically, 
given $\a=(a_1,\dots,a_m)$ and $\c_j=(c_{1j},\dots,c_{mj}) \in \IR^m$, $1\le j\le r$, there exist $A, C_1,\dots,C_r\in \cH_n$ such that 
$$A=\sum_{j=1}^rC_j \mbox{ with }\lambda(A)=\a, \mbox{ and } \lambda(C_j)=\c_j ,\ 1\le j\le r$$
if and only if $\tr A=  \sum_{j=1}^r\tr\(C_j \)$ and for every   $(J_0,J_1, \dots, J_r)\in LR(m,r)$, the inequality

\begin{equation}\label{ac}
 \sum_{i\in J_0}a_i\le \sum_{j=1}^r\sum_{i\in J_j}c_{ ij }  
\end{equation}
holds.  We have the following.

\begin{theorem} \label{3.1} Suppose $m \le n  $,   $\sigma_1\in \cD_m$ and $\sigma_2\in \cD_n$. The
following conditions are equivalent:
\begin{itemize}
\item[{\rm (1)}] There exists $\rho\in \cS(\sigma_1, \sigma_2)$ with rank $\le r$.
\item[{\rm (2)}] There exist   $C_1 ,\dots ,  C_r\in \cH_m$ and
$ \tilde C_1 ,\dots , \tilde C_r\in \cH_n$
such that
$$
{\rm (i)} \ \lambda(\tilde C_j)=\lambda(C_j\oplus O_{n-m}), \quad
{\rm (ii)} \  \sigma_1   = \sum_{j=1}^r C_j, \quad \hbox{ and } \quad
{\rm (iii)} \  \sigma_2    = \sum_{j=1}^r \tilde   C_j.$$
\item[{\rm (3)}] There exists $C\in \cH_m$ such that {\rm (2)}
holds with  $\lambda(C_j)=\lambda(C)$ and
$\lambda(\tilde C_j)=\lambda(C\oplus O_{n-m})$ for all $1\le j\le r$.
\end{itemize}
\end{theorem}

{\it Proof.} ``(1) $\Ra$ (2)'': Suppose there exists $\rho\in \cS(\sigma_1, \sigma_2)$ with rank $r$. Then there exist $W_1,\dots,W_r\in \cM_{m,n} $ such that
$$\sigma_1=W_1(W_1)^*+\cdots+W_r(W_r)^*\quad\mbox{ and }\quad\sigma_2=W^t_1\(W_1^t\)^*+\cdots+W^t_r\(W_r^t\)^*\,.$$

Then (2) holds with $C_j=W_jW_j^*$ and $\tilde C_j=W_j^t(W_j^t)^*$.

``(2) $\Ra$ (1)'' can be proven by reversing the above argument.

``(2) $\Ra$ (3)'': Suppose $\lambda(\sigma_1) = (a_1, \dots, a_m)$, and
$\lambda(C_j) = (c_{1j}, \dots, c_{mj})$ for $j = 1, \dots, r$.
Let $(J_0,J_1, \dots, J_r)\in LR(m,r)$, and
$\pi =(1,2,\dots,r)$ be the cyclic permutation of
$\{1,\dots,r\}$, i.e. $\pi(j) =j+1$ for $1\le j<r$ and $\pi(r)=1$. Then $(J_0,J_{\pi^k(1)}, \dots,
J_{\pi^k(r)})\in LR(m,r)$. Since $\sigma_1= C_{1} + \cdots + C_{r}$, by (\ref{ac})
 we have
$$
\begin{array}{rl}
&\sum_{i\in J_0}a_i\le \sum_{j=1}^r\sum_{i\in J_j}c_{i\sigma^k(j)}\
\mbox{ for every   }0\le k\le r-1\\&\\
 \Ra&\sum_{i\in J_0}a_i\le\sum_{j=1}^r\sum_{i\in J_j}c_{i},\
 \mbox{where } c_i=(c_{i1}+\cdots+c_{ir})/r\\&\\
 \Ra& \sigma_1 =\hat C_1 + \cdots + \hat C_r\end{array}
$$
for some  $\hat C_1 ,\dots , \hat C_r\in \cH_m$, where
$\lambda(\hat C_j)=(c_1,\dots,c_{m})$ for $1\le j\le r$. Similarly, we can choose $\tilde C_j$
such that $\lambda(\tilde C_j)=(c_1,\dots,c_{m},0,\dots,0)$.

Clearly, (3) $\Ra$ (2).
\qed

\begin{remark}  \rm  For every $m, n\ge 1$, there exists a  permutation matrix $P\in \cM_{mn}$ such that $ \rho  \in \cS(\sigma_1,\sigma_2)$ if and only if $P\rho P^T\in \cS(\sigma_2,\sigma_1)$. For example, let 
{\small
$$P=\[\begin{array}{cccccc}
1&0&0&0&0&0\\
0&0&0&1&0&0\\
0&1&0&0&0&0\\
0&0&0&0&1&0\\
0&0&1&0&0&0\\
0&0&0&0&0&1\end{array}\]\,.$$}
Then for every $\sigma_1\in \cD_2$ and $\sigma_1\in \cD_3$, $\rho\in\cS(\sigma_1,\sigma_2)$ if and only if $P\rho P^T\in \cS(\sigma_2,\sigma_1)$.  Thus, there is no loss of generality in the restriction of $m\le n$ in Theorem \ref{3.1}.

\end{remark}

\begin{example}\rm Suppose $m=2,\ n=3$, $\rho=\dfrac{1}{12}\left[
{\small \begin{array}{cccccc}
 3 & -2 & 1 & 1 & 2 & -1 \\
 -2 & 2 & 0 & 0 & -2 & 2 \\
 1 & 0 & 1 & 1 & 0 & 1 \\
 1 & 0 & 1 & 1 & 0 & 1 \\
 2 & -2 & 0 & 0 & 2 & -2 \\
 -1 & 2 & 1 & 1 & -2 & 3 \\
\end{array}}
\right]$. Then $\rho\in\cS(\sigma_1,\sigma_2)$ with
  $\sigma_1=\dfrac{1}{12}\left[
\begin{array}{cc}
 6 & -3 \\
 -3 & 6 \\
\end{array}
\right],\ \sigma_2=\dfrac{1}{12} \left[
\begin{array}{ccc}
 5 & 1 & 4 \\
 1 & 2 & 1 \\
 4 & 1 & 5 \\
\end{array}
\right]$. Let 
$$
C_1=\dfrac{1}{12}\left[
\begin{array}{cc}
 2 & 1 \\
 1 & 2 \\
\end{array}
\right],\  \
C_2=\dfrac{1}{12}\left[
\begin{array}{cc}
 4 & -4 \\
 -4 & 4 \\
\end{array}
\right],\  \
\tilde C_1=\dfrac{1}{12} \left[
\begin{array}{ccc}
 1 & 1 & 0 \\
 1 & 2 & 1 \\
 0 & 1 & 1 \\
\end{array}
\right],\ \ 
\tilde C_2=\dfrac{1}{12} \left[
\begin{array}{ccc}
 4 & 0 & 4 \\
 0 & 0 & 0 \\
 4 & 0 & 4 \\
\end{array}
\right].$$
We have 
\begin{itemize}
\item[{\rm (i)}] $\lambda(C_1)=\dfrac{1}{12}(3,1),\ \lambda(\tilde C_1)=\dfrac{1}{12}(3,1,0),\lambda(C_2)=\dfrac{1}{12}(8,0),\ \lambda(\tilde C_2)=\dfrac{1}{12}(8,0,0)$.
\item[{\rm (ii)}] $\sigma_1=C_1+C_2$.
\item[{\rm (iii)}]  $\sigma_2=\tilde C_1+\tilde C_2$
\end{itemize}

Let $c_1=\dfrac{1}{2}\(\dfrac{3}{12}+\dfrac{8}{12}\)=\dfrac{11}{24}$ and $c_2=\dfrac{1}{2}\(\dfrac{1}{12}+0\)=\dfrac{1}{24}$. Then we can choose $$
C'_1=\dfrac{1}{24}\left[
\begin{array}{cc}
 2 & -3 \\
 -3 & 10 \\
\end{array}
\right],\ 
C_2=\dfrac{1}{24}\left[
\begin{array}{cc}
 10 & -3 \\
 -3 & 2 \\
\end{array}
\right],\ 
$$ 
and 
$$
\hat  C_1=\dfrac{1}{24} \left[
{\small \begin{array}{ccc}
 5+4 \sqrt{\frac{2}{57}} & 1-14 \sqrt{\frac{2}{57}} & 4+4 \sqrt{\frac{2}{57}} \\
 1-14 \sqrt{\frac{2}{57}} & 2-8 \sqrt{\frac{2}{57}} & 1-14 \sqrt{\frac{2}{57}} \\
 4+4 \sqrt{\frac{2}{57}} & 1-14 \sqrt{\frac{2}{57}} & 5+4 \sqrt{\frac{2}{57}} \\
\end{array}}
\right],\ 
\hat C_2=\dfrac{1}{24} \left[
{\small \begin{array}{ccc}
 5-4 \sqrt{\frac{2}{57}} & 1+14 \sqrt{\frac{2}{57}} & 4-4 \sqrt{\frac{2}{57}} \\
 1+14 \sqrt{\frac{2}{57}} & 2+8 \sqrt{\frac{2}{57}} & 1+14 \sqrt{\frac{2}{57}} \\
 4-4 \sqrt{\frac{2}{57}} & 1+14 \sqrt{\frac{2}{57}} & 5-4 \sqrt{\frac{2}{57}} \\
\end{array}}
\right].$$

Then we have 
$\sigma_1=C'_1+C'_2$, $\sigma_2=\hat C_1+\hat C_2$ with 
$\lambda(C'_1)=\lambda(C'_2)=(c_1,c_2)$ and $\lambda(\hat C_1)=\lambda(\hat C_2)=(c_1,c_2,0)$.

\end{example}

When $n=rm$, we have the following corollary of Theorem \ref{3.1}.

\begin{corollary}\label{cor}  Suppose $  n=rm  $,   $\sigma_1\in \cD_m$ and $\sigma_2\in \cD_n$. The
following conditions are equivalent:
\begin{itemize}
\item[{\rm (a)}] There exists $\rho\in \cS(\sigma_1, \sigma_2)$ with rank $\le r$.
\item[{\rm (b)}] There exists $C=(C_{ij})\in H_n $ with $C_{ij}\in M_m$ such that  $\lambda(C)=\lambda(\sigma_2)$ and $\lambda \( C_{11}+\cdots +C_{rr}  \) =\lambda(\sigma_1)$.

\item[{\rm (c)}] Condition
{\rm (b)} holds with  $\lambda\(C_{11}\)=\cdots =\lambda\(C_{rr}\)  $.
\end{itemize}
\end{corollary}

{\it Proof.} (a) $\Ra $ (b):  Suppose (a) holds. Let  $C_j$ and $\tilde C_j$, $1\le j\le r$,  be as given by condition (2) in Theorem \ref{3.1}. For each $1\le j\le r$, there exists an $n\times n$ unitary matrix $U_j$ such that $\tilde C_j=~U_j\(C_j\oplus  O_{(r-1)m}\)U_j^* $. Let $V_j\in \cM_{n,m}$ be formed by the first $m$ columns of $U_j$. Let  $R_j= V_jC_j^{1/2}$ and $R=\[R_1|\cdots |R_r\]$. Set $C=R^*R$. Then 

$$\lambda(C)=\lambda(R^*R)=\lambda(RR^*)=\lambda\(\sum_{j=1}^r\tilde C_j\)=
\lambda(\sigma_2) $$
and 
$$\lambda \( C_{11}+\cdots +C_{rr}  \) =\lambda\(\sum_{j=1}^r R_j^*R_j\)= \lambda\(\sum_{j=1}^r C_j\) = \lambda(\sigma_1)\,.$$
Therefore, $C=R^*R$ will satisfy condition (b).

(b) $\Ra $ (a): Suppose (b) holds. Let $C=R^*R$ where $R=\[R_1|\cdots |R_r\]$, with $R_j\in \cM_{n,m}$. Then $C_{ii}=R^*_iR_i$ and $\lambda(C)=\lambda(RR^*)$. Hence, we have

$$\lambda(\sigma_1)=\lambda\(R_1^*R_1+\cdots +R_r^*R_r\)\mbox{ and }  \lambda(\sigma_2)= \lambda\(R_1^*R_1+\cdots +R_r^*R_r\)\,.$$
Since $\lambda( R_i^*R_i)= \lambda \(R_iR_i^*\oplus O_{(r-1)m}\)$, the result follows from condition (2) in  Theorem \ref{3.1}.

 (b) $\Lra$ (c) follows from (2)  $\Lra$ (3) in Theorem \ref{3.1}
 \qed

Applying Theorem \ref{3.1}, we have the following algorithm 
for finding the minimal rank $r$ of matrices in $\cS(\sigma_1, \sigma_2)$.

\medskip
\noindent\fbox{\parbox{\textwidth}{
\begin{algorithm} \label{alg2.5}\rm
Suppose $\sigma_1\in \cD_m$ and   $\sigma_2\in \cD_n$, with $n\ge m$. Let
$$\lambda(\sigma_1)=(a_1,\cdots,a_m)\quad
\hbox{ and } \quad \lambda(\sigma_2)=(b_1,\cdots,b_n).$$

\medskip
{\bf Step 1} If $b_i=a_i$ for $1\le i\le m$, then $r=1$. If not, $r>1$, then go to step 2.

\medskip
{\bf Step 2} For $r>1$, suppose the previous steps shows that the minimal rank is $\ge r$.
Let
$$\begin{array}{rl} P_r(a_1,\cdots,a_m)=\left\{\c\in\IR^m_{\downarrow}:
\right.& c_1, \dots, c_m \ge 0,
\sum_{j=1}^mc_j=1/r,  \\
&\left.\sum_{i\in J_0}a_i\le\sum_{j=1}^r\sum_{i\in J_j}c_{i} \
\mbox{ for all }(J_0,J_1, \dots, J_r)\in LR(m,r)\right\},\\&\\
Q_r(b_1,\cdots,b_n)=\left\{\c\in\IR^m_{\downarrow}:
\right.& c_1, \dots, c_m \ge 0,\sum_{j=1}^mc_j=1/r, \\
&\left.\sum_{i\in J_0}b_i\le\sum_{j=1}^r\sum_{i\in J_j}\hat c_{i} \  \mbox{ for all }
(J_0,J_1, \dots, J_r)\in LR(n,r)\right\},
\end{array}$$
where  in $Q_r(b_1, \dots, b_n)$,
$\hat c_i=c_i$ for $1\le i\le m$ and $\hat c_i=0 $ for $m+1\le i\le n$.

\medskip
If $S=P_r(a_1,\cdots,a_m)\cap Q_r(b_1,\cdots,b_n)$ is non-empty, then
the minimal rank of $\rho$ in $\cS(\sigma_1, \sigma_2)$ is $ r$.

Otherwise, the minimal rank is larger than  $r$ and we have to repeat Step 2 with $r$ increased by 1.
\end{algorithm}
}
}
\medskip

By Theorem \ref{thm2.4}, Algorithm \ref{alg2.5} will terminate for some
$r \le \max\{\rank(\sigma_1), \rank(\sigma_2)\}$.
In fact, we can focus on $\sigma_1\in \cD_m$ and $\sigma_2\in \cD_n$ with rank
$m$ and $n$, respectively.

The set $P_r(a_1,\cdots,a_m)$ and $Q_r(b_1,\cdots,b_n)$ are polyhedral.
There are standard linear programming packages for checking whether
$S=P_r(a_1,\cdots,a_m)\cap Q_r(b_1,\cdots,b_n)$ is empty.
Actually, we have the following.

\begin{proposition} Let
$S=P_r(a_1,\cdots,a_m)\cap Q_r(b_1,\cdots,b_n)$ be defined as in Algorithm
{\rm \ref{alg2.5}}. Then the set $S$ is non-empty
if and only if $a_1, \dots, a_m, b_1, \dots, b_n$
satisfy a finite set of linear inequalities.
\end{proposition}

\it Proof. \rm
Because $(a_1/r, \dots, a_m/r) \in  P_r(a_1, \dots, a_m)$,
and $P_r(a_1, \dots, a_m)$ is governed by a finite set of inequalities,
it is a non-empty polyhedron. Thus, there are finitely many extreme points
expressed as linear combinations of $a_1, \dots, a_m$.
Now, $Q_r(b_1, \dots, b_n)$ is determined a finite set of inequalities, say,
$v_j^t x \le \beta_j$ for $j = 1, \dots, N$,
where $v_1, \dots, v_N \in \IR^n$ with entries in
$\{b_1, \dots, b_n\}$ and $\beta_1, \dots, \beta_N \in \IR$.
If all the extreme points of $P_r(a_1, \dots, a_m)$ lies in the complement
of the half space defined by $v_1^t x \le \beta_1$, then $S$ is empty.
Otherwise, the intersection of $P_r(a_1, \dots, a_m)$ and the half space
defined by $v_1^t x \le \beta_1$ is a non-empty polytope, and has a finite
number of extreme points expressed as linear combinations of $a_1, \dots, a_m, b_1, \dots, b_n$.
We can repeat the argument to this new polytope and the half space $v_2^t x \le \beta_2$.
We may conclude either the set and the half space has empty intersection or non-empty
intersection. Repeating the process, we get a finite set of inequalities involving
$a_1, \dots, a_m, b_1, \dots, b_n$, such that any one of them being violated will
imply that $S = \emptyset$, and $S \ne \emptyset$ if all the inequalities are satisfied.
\qed

By the above proposition, one can determine whether
$S = P_r(a_1,\cdots,a_m)\cap Q_r(b_1,\cdots,b_n)$ is non-empty by checking a
finite set of inequalities in terms of $a_1, \dots, a_m, b_1, \dots, b_n$.
Using this result, one may determine the set
$$\cS_r(m: \sigma_2)= \{\sigma \in \cD_m: \hbox{ there is }
\rho \in \cS(\sigma, \sigma_2) \mbox{ with rank at most }\ r \}$$
for a given  $\sigma_2\in \cD_n$; and
$$\cS_r(\sigma_1  :n)= \{\sigma  \in \cD_n: \hbox{ there is }
\rho \in \cS(\sigma_1, \sigma) \mbox{ with rank at most }\ r \}$$
for a given  $\sigma_1\in \cD_m$. We have the following.

\begin{proposition} Suppose $\sigma_2 \in \cD_n$ has eigenvalues $b_1 \ge \cdots \ge b_n$.
Then $\sigma \in S_r(m: \sigma_2)$ if and only if $\sigma$ has eigenvalues
$a_1 \ge \cdots \ge a_m$ such that $P_r(a_1, \dots, a_m) \cap Q_r(b_1, \dots, b_n)
\ne \emptyset$.

Suppose  $\sigma_1 \in \cD_m$ has eigenvalues $a_1 \ge \cdots \ge a_m$.
Then $\sigma \in S_r(\sigma_1 :n)$ if and only if $\sigma$ has eigenvalues
$b_1 \ge \cdots \ge b_m$ such that $P_r(a_1, \dots, a_m) \cap Q_r(b_1, \dots, b_n)
\ne \emptyset$.
\end{proposition}

Although one can determine whether the set
$S = P_r(a_1,\cdots,a_m)\cap Q_r(b_1,\cdots,b_n)$ is non-empty by checking a
finite set of inequalities,
the number of inequalities involved may be very large.
For low dimension case, the inequalities may be reduced to a smaller set
after the redundant inequalities are removed. We illustrate this in the
following proposition with a direct proof. It would be nice if one can give a
description of non-redundant inequalities governing the eigenvalues of  
the reduced states of a bipartite state with prescribed rank.

Given $\a=(a_1, \dots,a_n)$ and $\b=(b_1, \dots,b_n)\in \IR^n$, we say that $\a$ is majorized by $\b$, denoted by $\a\prec\b$, if  for $1\le k\le n-1$ the sum of the $k$ largest components of $\a$ is less than or equal to that of $\b$. By Horn's result \cite{H}, $\a$ is the diagonal of  some $B\in H_n$ with eigenvalues $b_1, \dots,b_n$ if and only if $\a\prec \b$.

\begin{proposition} \label{prop2.7}  Suppose  $\sigma_1 \in \cD_3$ has eigenvalues
$a_1 \ge a_2 \ge a_3$   and  $\sigma_2 \in \cD_6$ has eigenvalues
$b_1 \ge \cdots \ge b_6$. Then $\sigma_1 \in \cS_2(3:\sigma_2)$
if and only if   $a_1 , a_2 , a_3$ satisfying $\sum_{i=1}^3a_i= \sum_{j=1}^6b_j$ and the following
inequalities:
\begin{equation}\label{ab3}\begin{array}{rcl}
b_3+b_6,\ b_4+b_5&\le a_1\le &b_1+b_2\\&&\\
\dfrac{b_3+b_4+b_5+b_6}2&\le a_2\le &\dfrac{b_1+b_2+b_3+b_4}2\\&&\\
b_5+b_6&\le a_3\le &b_1+b_4,\ b_2+b_3\end{array}\end{equation}
\end{proposition}


\it Proof. \rm Suppose $\sigma_2 \in \cD_6$ has eigenvalues
$b_1 \ge \cdots \ge b_6$ and  $\sigma_1 \in \cS_2(3:\sigma_2)$ has eigenvalues $a_1 \ge a_2 \ge a_3$. Then by Corollary \ref{cor}, there exists a unitary matrix $U\in M_6$ such that $U^*\sigma_2 U=\(C_{ij}\)_{1 \le i, j \le 2}$ with
$$ \lambda(C_{11})=\lambda(C_{22})=(c_1, c_2, c_3) \ \mbox{ and }
\lambda(C_{11}+C_{22})=  (a_1,a_2,a_3)$$

Then there exist $c_1 \ge c_2 \ge c_3$ and $3\times 3$ unitary matrices $V_1 $ and $V_2$ such that $\diag\(V_i^*C_{ii}V_i\)=(c_1, c_2, c_3)$ for $i=1,\ 2$. Hence,   $(c_1,c_1,c_2,c_2,c_3,c_3)\prec (b_1, \cdots, b_6)$ \cite{H} and  we have
\begin{enumerate}
\item $a_1\le 2c_1\le b_1+b_2$.
\item $a_2\le c_1+c_2\le \dfrac{b_1+b_2+b_3+b_4}2$.
\item $a_3\le 2c_2=(b_1+b_2+b_3+b_4+b_5+b_6)-2(c_1+c_3)$

\quad $\le (b_1+b_2+b_3+b_4+b_5+b_6)-(b_2+b_3+b_5+b_6)= b_1 + b_4,$ \quad and

\quad $(b_1+b_2+b_3+b_4+b_5+b_6)-(b_1+b_4 +b_5+b_6)=b_2+b_3\,.$
\end{enumerate}
Here,    $2(c_1+c_3)\ge  b_2+b_3+b_5+b_6,\ b_1+b_4 +b_5+b_6$
follows from the fact that \cite{Fu}  $$(\{2,3,6,7\},\ \{1,3,4,5\},\ \{1,3,4,5\}),\ (\{1,4,5,6\},\ \{1,3,4,5\},\ \{1,3,4,5\})\in LR(6,2)\,.$$
The other inequalities can be deduced by looking at 
$2\mu I_3-\diag(a_1,a_2,a_3)$ and $\mu I_6-\diag(b_1,  \dots, b_6) $, where
$\mu = (b_1 + \cdots + b_6)/6$.

 Given  $b_1 \ge \cdots \ge b_6$, let $S$ be the set of $(a_1,a_2,a_3)$ satisfying
$a_1\ge a_2\ge a_3$, $\sum_{i=1}^3a_i
=\sum_{i=1}^6b_i=6\mu$ and  (\ref{ab3}).
 Then $S$
is a convex polyhedron in $\IR^3$. If $S$ is non-empty, we can choose an extreme point $(a_1,a_2, a_3)$ of $S$.  Therefore, among the inequalities $a_1\ge a_2\ge a_3$ and  (\ref{ab3}),  at least two equalities hold. If $a_1=a_2=a_3= 2\mu$,
then from (\ref{ab3}) we have
$$\begin{array}
{rcl}
b_3+b_6,\ b_4+b_5&\le 2\mu\le &b_1+b_4,\ b_2+b_3\\&&\\
b_5 \le 2\mu - b_4 \le b_1   &\hbox{ and }&
b_6 \le 2\mu - b_3 \le b_2\end{array}.$$
Thus, $\diag(b_1, b_5)$ is unitarily similar to $B_1$ with diagonal
entries $2\mu -b_4, b$ and $\diag(b_2, b_6)$ is unitarily similar to
$B_2$ with diagonal entries $2\mu - b_3, c$; see \cite{H}.
Thus, $B = \diag(b_1, \dots, b_6)$ is unitarily similar to
$\diag(b_3,b_4) \oplus B_1 \oplus B_2$. There exists a permutation matrix $P$ such that
$P^*BP=(D_{ij})$
with $D_{11} = \diag(b_3, 2\mu - b_4, b)$
and $D_{22} = \diag(2\mu-b_3, b_4, c)$.
By the trace condition, we see that $b+c = 2\mu$. The result follows from Corollary \ref{cor}.

\medskip
Suppose either $a_1>a_2$ or $a_2>a_3$. Then at least one of the equalities in (\ref{ab3}) holds.
 Consider the following cases:
\begin{enumerate}
\item  $a_1= b_1+b_2$. Then we have $\dfrac{b_3+b_4+b_5+b_6}2 \le a_2=(b_3+b_4+b_5+b_6)-a_3
\le b_3+b_4$. Therefore, $(a_2,a_3)\prec (b_3+b_4,b_5+b_6)$. So there exists a $2\times 2$ unitary matrix $U_1$ such that $U_1^*\diag(b_3+b_4,b_5+b_6)U_1=(a_2,a_3)$. Let $U_2=U_1\diag(1,-1)$. Let $$B_1=\dfrac{1}{2}\[\begin{array}{cc}
b_3+b_4&b_3-b_4\\
b_3-b_4&b_3+b_4\end{array}\]\mbox{ and }B_2=\dfrac{1}{2}\[\begin{array}{cc}
b_5+b_6&b_5-b_6\\
b_5-b_6&b_5+b_6\end{array}\].$$ Then $B=\diag(b_1,b_2)\oplus B_1\oplus B_2$ has eigenvalues $b_1 , \dots , b_6$.  There exists a permutation matrix $P$ such that
$P^*BP=(D_{ij})$
with $$D_{11} = \dfrac{1}{2}\diag(2b_1, b_3+b_4,b_5+b_6)\mbox{
and }D_{22} = \dfrac{1}{2}\diag(2b_2,b_3+b_4,b_5+b_6).$$ Let $U=P\([1]\oplus U_1\oplus [1]\oplus U_2\)$. Then $U^*BU$ will satisfy   condition (2) in Corollary \ref{cor}.

  \item $a_2=\dfrac{b_1+b_2+b_3+b_4}2$.
  Then $a_1+a_2\ge 2a_2=b_1+b_2+b_3+b_4\Ra a_3\le b_5+b_6$. Therefore, $ a_3= b_5+b_6$ and $(a_1,a_2)\prec (b_1+b_2,b_3+b_4)$.
  Thus, the result follows as in the previous case.
\item $a_3=b_1+b_4$. Then$$\begin{array}{rl}
      & (b_1+b_2+b_3+b_4+b_5+b_6)=a_1+a_2+a_3\ge 3a_3 \\&\\
      & =3(b_1+b_4) \ge (b_1+b_2+b_3+b_4+b_5+b_6)\\&\\
      \Ra&a_1=a_2=a_3=b_1+b_4=b_2+b_5=b_3+b_6\\&\\
      \Ra&C=\diag(b_1,\dots,b_6)\mbox{ will satisfy (2) in Corollary \ref{cor}}\end{array}$$
\item $a_3=b_2+b_3$. Then $a_1+a_2=b_1+b_4+b_5+b_6$ and $a_2\ge a_3\ge b_5+b_6$. Therefore,
$(a_1,a_2)\prec (b_1+b_4,b_5+b_6)$.
  Thus, the result follows as in the   Case 2.
\end{enumerate}
The proof for the other  equalities are similar. \qed

\medskip
Note that the same set of inequalities (\ref{ab3})
will determine whether $\sigma \in \cD_6$ with eigenvalues
$b_1 \ge \cdots \ge b_6$ lying in $\cS_2(\sigma_1:6)$ for a given $\sigma_1\in \cD_3$ with
eigenvalues $a_1 \ge a_2 \ge a_3$.

In case $a_3 = 0$, then $b_5 = b_6 = 0$, and the set of inequalities reduce to:
$$(b_3+b_4)/2 \le a_2, \qquad \hbox{ and } \qquad a_1 \le b_1+b_2.$$
These inequalities will determine
$\sigma \in \cS_2(\sigma_1:4)$ with eigenvalues $b_1 \ge \cdots \ge b_4$
for a given $\sigma_1 \in \cD_2$ with eigenvalues $a_1 \ge a_2$. The same set of inequalities
will also determine $\sigma\in \cS_2(2:\sigma_2)$ with eigenvalues $a_1 \ge a_2$
for a given $\sigma_2 \in \cD_4$ with eigenvalues $b_1 \ge \cdots \ge b_4$.

Note that
$\a$ satisfies (\ref{ab3}) if and only if $(c_1,c_2,c_3)\prec \a\prec (b_1+b_2, b_3+b_4,b_5+b_6)$, where
$$\c=\left\{\begin{array}{ll}
\(b_4+b_5,\dfrac{b_1+b_2+b_3+b_6}2,\dfrac{b_1+b_2+b_3+b_6}2\)&\mbox{ if }    \dfrac{1}{3}\le b_4+b_5\\&\\
(\dfrac{1}{3},\dfrac{1}{3},\dfrac{1}{3})&\mbox{ if }b_4+b_5\le  \dfrac{1}{3}\le b_2+b_3\\&\\
\( \dfrac{b_1+b_4+b_5+b_6}2,\dfrac{b_1+b_4+b_5+b_6}2,b_2+b_3\)&\mbox{ if }     b_2+b_3 \le \dfrac{1}{3}\,. \end{array}\right.$$

\section{Quantum channels}

Recall that quantum channels are completely positive linear maps
$\Phi: \cM_m \rightarrow \cM_n$ that admit the operators sum representation
\begin{equation}\label{operatorsum}
\phi(A)=\sum_{j=1}^rF_jAF_j^*,
\end{equation}
for some $n\times m$ matrices $F_1, \dots, F_j$ such that
$\sum_{j=1}^r F_j^*F_j = I_m$; see \cite{C,K}.
By the result in \cite{C},
$\Phi$ is a quantum channel if and only if the Choi matrix
$C(\Phi) = (\Phi(E_{ij})) \in \cM_m(\cM_n)$
is positive semidefinite and $\tr \,\Phi(E_{ij}) = \delta_{ij}$.
Thus, the set of quantum channels can be identified with the set
\begin{eqnarray*}
QC(m,n) &=&
\{P = (P_{ij}) \in \cM_m(\cM_n): P \hbox{ is positive semidefinite},
(\tr(P_{ij})) = I_m\} \\
&=& \{ m\rho \in \cD_{mn}: \tr_2(\rho) = I_m/m\}.
\end{eqnarray*}
Consequently, the set of quantum channels
$\Phi: M_m \rightarrow M_n$ satisfying $\Phi(I_m/m) = \rho_2\in \cD_n$
can be identified with $\cS(I_m/m, \rho_2)$.
In particular,
$\cS(I_n/n,I_n/n)$ can be identified with the set of unital
quantum channels from $M_n$ to $M_n$.

For a quantum channel $\Phi$, its Choi
rank is defined as the rank of its Choi matrix $C(\Phi)$. Moreover, it is known that
$\Phi$ has Choi rank $r$ if and only if $r$ is the minimum number of matrices
$F_1, \dots, F_r$  required in the operator sum representation of $\Phi$.
By Theorem \ref{cor2.3}, we have the following.

\begin{proposition}
There is $\rho \in \cS(I_n/n,I_n/n)$ of rank $k$ if and only if $1 \le k \le n^2$.
Equivalently, there is a unital quantum channel with Choi rank $k$ if and only if
$1 \le k \le n^2$.
\end{proposition}

By the result in the previous section, we have the following.

\begin{proposition}
Let $\rho_2 \in \cD_n$. There is a quantum channel $\Phi: \cM_m \rightarrow \cM_n$
with Choi rank $k$ and $\Phi(I_m/m) = \rho_2$ if and only if there is a
rank $k$ element in $\cS(I_m/m, \rho_2)$. As a result, the value $k$ can be any
value between
the minimum value $r$ determined by Algorithm {\rm \ref{alg2.5}}  and the maximum
value $\rank(\rho)\cdot m$.
\end{proposition}

\bibliographystyle{amsplain}


\section*{Acknowledgment}

We thank Dr.\ Cheng Guo for some helpful discussion. Also, we would like
to thank the two referees for some helpful suggestions.
Li is an honorary professor of the Shanghai University,
and an affiliate member of the Institute for Quantum Computing, University of
Waterloo; his research was supported by
the USA NSF DMS 1331021, the Simons Foundation
Grant 351047, and NNSF of China Grant 11571220.
Part of the work
was done while the Li and Poon were visiting the Institute for Quantum Computing
at the University of Waterloo.
They gratefully acknowledged the support and kind hospitality of the Institute.

\end{document}